\documentclass[apl,aps,amssymb,twocolumn,showpacs,floatfix]{revtex4}

\usepackage{graphicx}% Include figure files
\usepackage{dcolumn}% Align table columns on decimal point
\usepackage{bm}% bold math

\begin{document}

\preprint{}

\title{Optical and electrical spin injection and spin transport in hybrid Fe/GaAs devices}

\author{S. A. Crooker, M. Furis}

\affiliation{National High Magnetic Field Laboratory, Los Alamos
National Laboratory, Los Alamos, NM 87545}

\author{X. Lou, P. A. Crowell}

\affiliation{School of Physics and Astronomy, University of
Minnesota, 116 Church St. SE, Minneapolis, MN 55455}

\author{D. L. Smith}

\affiliation{Theoretical Division, Los Alamos National Laboratory,
Los Alamos, NM 87545}

\author{C. Adelmann, C. J. Palmstr{\o}m}

\affiliation{Department of Chemical Engineering and Material
Science, University of Minnesota, 421 Washington Ave. SE,
Minneapolis, MN 55455}

\date{\today}
\begin{abstract}

We discuss methods for imaging the nonequilibrium spin polarization of
electrons in Fe/GaAs spin transport devices.  Both optically- and
electrically-injected spin distributions are studied by scanning
magneto-optical Kerr rotation microscopy. Related methods are used
to demonstrate electrical spin detection of optically-injected
spin polarized currents. Dynamical properties of spin transport are
inferred from studies based on the Hanle effect, and the influence
of strain on spin transport data in these devices is discussed.
\end{abstract}

%\pacs{}
\maketitle
%---------------------------------------------------------------
The demonstration of electrical spin injection and spin detection in
lateral metallic devices, including spin-valve and spin precession
effects \cite{Johnson, Jedema, Valenzuela}, has generated
considerable interest in related devices based on semiconductors.
Unlike their metallic counterparts, characterization of these
`semiconductor spintronic' structures benefits from the many
magneto-optical tools that have been developed over the years to
probe spin-polarized electrons and holes in semiconductors
\cite{OO}.  In this paper we describe experiments that measure and image both
optically- and electrically-injected spin polarizations in GaAs
using scanning magneto-optical Kerr rotation microscopy. These
techniques are applied to hybrid Fe/GaAs lateral spin transport
structures. Using cw lasers and small magnetic fields to induce
electron spin precession, dynamical properties of spin transport are
inferred from Hanle-effect studies and theoretical models of the
spin drift-diffusion equations. The influence of strain on spin
transport measurements is also discussed. Related techniques are
used to demonstrate electrical spin detection of optically-injected
spin polarized currents in these devices.

Figure 1 shows a schematic of the experiment. The Fe/GaAs devices
are mounted, nominally strain-free, on the variable-temperature cold
finger of a small optical cryostat (all presented data were acquired
at 4 K). The cryostat itself is mounted on a \emph{x-y} stage. The
samples may also be held by a small cryogenic vise machined into the
cold finger \cite{CrookerPRL}. The uniaxial stress applied to the
sample by the vise is uniform and can be varied \emph{in situ} by a
retractable actuator.  For devices grown on [001] oriented GaAs
substrates and cleaved along the usual $\langle110\rangle$ crystal
axes, this uniaxial (shear) stress leads to nonzero off-diagonal
elements of the crystallographic strain tensor in GaAs,
$\epsilon_{xy}$.  $\epsilon_{xy}$ couples directly to electron spin
($\sigma$) and momentum (\textbf{k}) via spin-orbit coupling, leading to effective magnetic
fields `seen' by moving electrons \cite{OO, CrookerPRL}.

\begin{figure}[tbp]
\includegraphics[width=.43\textwidth]{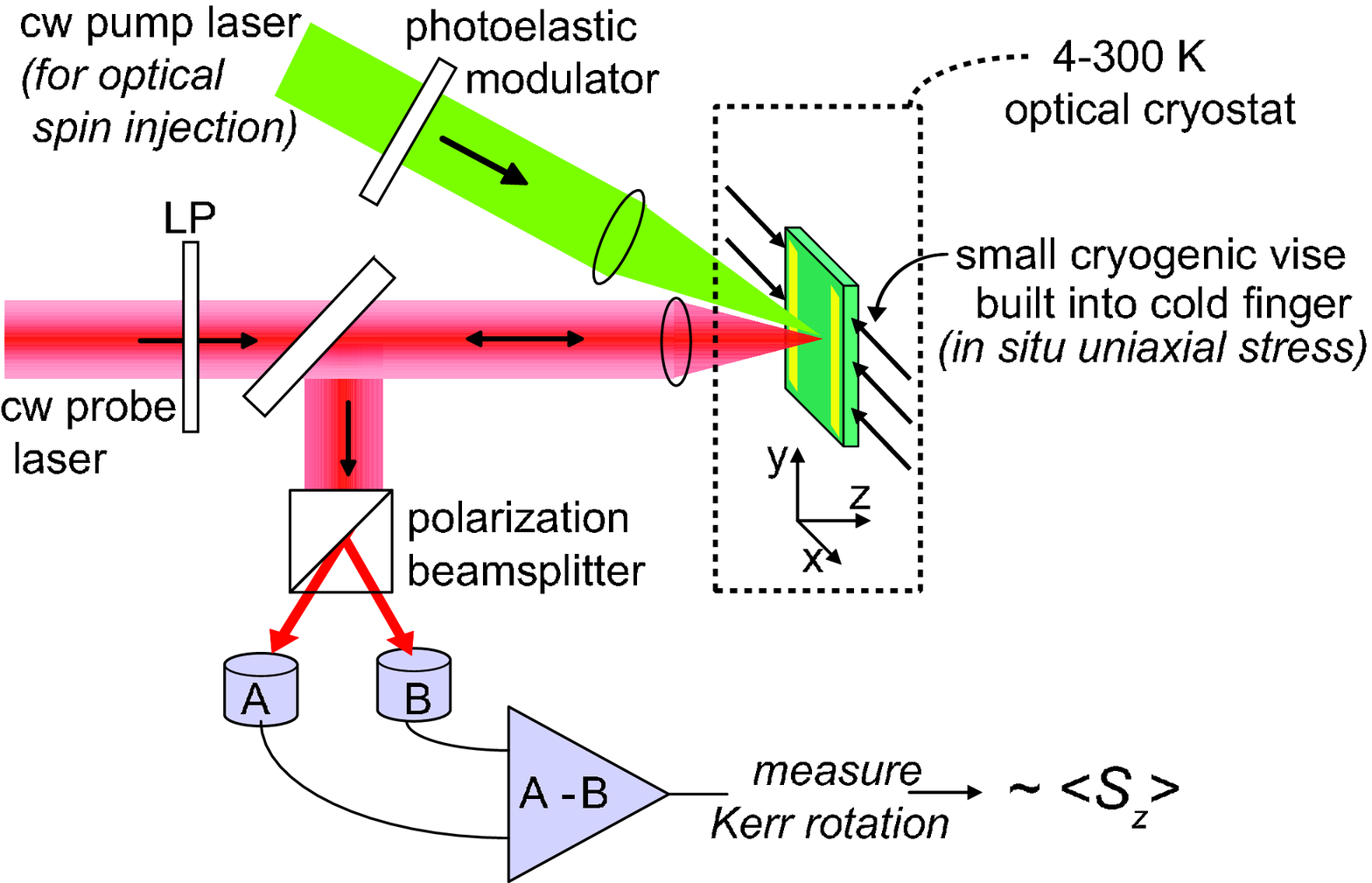}
\caption{(Color online)  A schematic of the scanning Kerr microscope
used to image optically- and/or electrically-injected electron spins
in Fe/GaAs devices. The measured polar Kerr rotation imparted on the
reflected probe laser beam is proportional to the out-of-plane
($\hat{z}$) component of the conduction electron spin polarization,
$S_z$. External coils (not drawn) control the applied
magnetic fields $B_x, B_y, B_z$.} \label{fig1}
\end{figure}

The steady-state spin polarization of conduction electrons in the GaAs is
measured by the polar magneto-optical Kerr effect. As has been
briefly described in recent works \cite{CrookerPRL, CrookerScience,
Hruska, Furis}, a cw probe laser beam, derived from a narrowband and
frequency tunable Ti:sapphire ring laser, is linearly polarized and
focused tightly to a 4 $\mu$m spot on the sample. The Kerr rotation
(\emph{i.e.}, optical polarization rotation) imparted to the
reflected probe laser is proportional to the out-of-plane
($\hat{z}$) component of electron spin, $S_z$. This Kerr rotation
(KR) is measured by balanced photodiodes using lock-in techniques.
To measure optically-injected spins, a 1.58 eV cw pump laser is also
focused to a 4 $\mu$m spot on the device.  The polarization of this
pump laser is modulated from left- to right- circular (injecting
spins oriented along $\pm \hat{z}$) by a 50 kHz photoelastic
modulator.  To measure electrically-injected spins, the electrical
bias applied to the Fe contacts is square-wave modulated at 3.1 kHz.
The cryostat and/or the probe laser can be raster-scanned in the
\emph{x-y} plane to acquire a 2D image of the electron spin
polarization $S_z$.  We simultaneously image the reflected probe
intensity to infer the topography of the device surface. The applied magnetic field is controlled by external coils.

\begin{figure}[tbp]
\includegraphics[width=.40\textwidth]{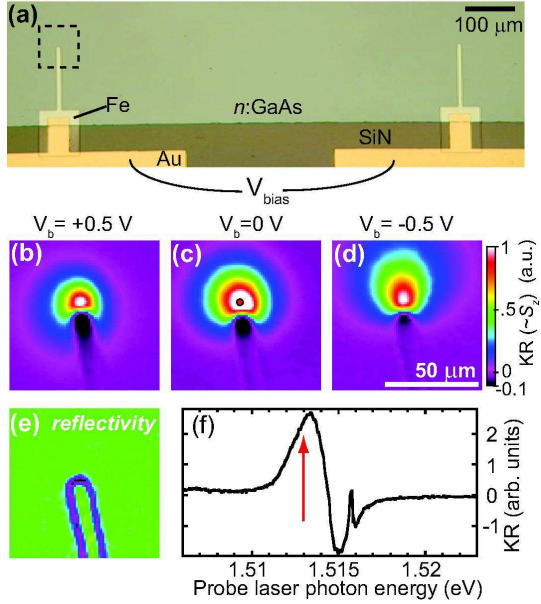}
\caption{(Color online)(a) A photomicrograph of a lateral Fe/GaAs device
having 10 $\mu$m wide Fe ``fingers" on a \emph{n}:GaAs epilayer. The
dotted square shows the $80 \times 80$ $\mu$m imaged region. (b-d)
Kerr rotation (KR) images of electron spin polarization $S_z$ that
was \emph{optically} injected into the \emph{n}:GaAs just off the
tip of a Fe finger. The red dot shows the 4 $\mu$m injection spot.
The dc electrical bias was $V_b$ = +0.5, 0, and -0.5 V respectively
($I=600$ $\mu$A at $V_b=+0.5$ V). These spins are seen to flow into
(away from) the Fe finger at positive (negative) bias. (e) An image
of the reflected probe power, used to infer topographical features.
(f) The measured KR due to by optically-injected spins in this
device, versus photon energy of the probe laser.  This spin-dependent
spectral `fingerprint' is different for every sample (arrow
indicates the probe energy used to acquire the images).}
\label{fig2}
\end{figure}

Figure 2(a) shows a photomicrograph of a Fe/GaAs ``finger"
device. All devices were fabricated from Fe/GaAs heterostructures grown by molecular beam epitaxy as
described in Refs. \cite{CrookerScience, XiaohuaPRL}.  Briefly: on
(001) oriented semi-insulating GaAs, 300 nm of undoped GaAs was grown, followed by a 2 $\mu$m epilayer of Si-doped
\emph{n}:GaAs having electron doping in the range $n=1-5
\times 10^{16}$/cm$^3$ to maximize the low-temperature electron spin
lifetime and spin transport length \cite{Kikkawa,DzhioevPRB,Furis}.
Then a 15 nm layer was grown where the doping was rapidly increased
to $n^+=5 \times 10^{18}$/cm$^3$, followed by a 15 nm layer doped
uniformly at $n^+=5 \times 10^{18}$/cm$^3$. These heavily-doped
layers define a narrow Schottky barrier through which electrons
can tunnel \cite{Hanbicki}. Then 5 nm of Fe was epitaxially
deposited, followed by 2 nm of Al. To define the lateral
structures, the metal and $n^+$:GaAs were etched away except for the Fe contact regions. Gold contacts were
deposited after a SiN insulation layer.

\begin{figure}[tbp]
\includegraphics[width=.40\textwidth]{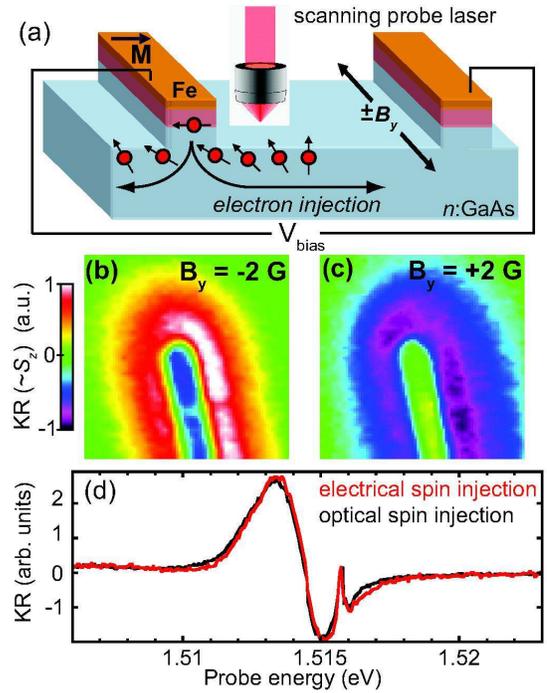}
\caption{(Color online) Imaging electrical spin injection. (a) Cartoon of a Fe/GaAs heterostructure in cross-section. Electrons
tunnel from Fe into the \emph{n}:GaAs with initial spin polarization
$S_0$ antiparallel to \textbf{M} (in this drawing). A small
orthogonal magnetic field $\pm B_y$ is used to precess the injected
spins out-of-plane (along $\pm \hat{z}$) so that they can be
measured by the polar Kerr effect. (b) $80 \times 80$ $\mu$m image
of electrically injected spin polarization ($V_b=-0.5$ V, $I=600$
$\mu$A, $B_y=-2$ G). (c) The same, but with $B_y$ reversed so that
spins precess into the page giving negative KR signal. (d) KR vs
probe laser photon energy for this case of electrically-injected
spin polarization (red line), showing good agreement with the prior
case of optically-injected spins (black).} \label{fig3}
\end{figure}

The images in Figs. 2(b-d) show the drift and diffusion of
\emph{optically} injected electron spin polarization in the vicinity of one
Fe finger (dotted region in Fig. 2(a)).  Spins are optically
injected just off the tip of the left Fe finger, and the dc
electrical bias applied to this Fe finger (relative to the rightmost
Fe finger) is $V_b = +0.5$, 0, and -0.5 V respectively. At zero
bias, the spins diffuse radially away from the point of injection
with a spin diffusion length of order 10 $\mu$m.  At positive
(negative) bias, these optically injected spins are directly
observed to flow into (away from) the Fe finger, which is acting as
a drain (source) of electron current. The reflectivity image of Fig.
2(e) shows the border of the Fe finger.

\begin{figure*}[tbp]
\includegraphics[width=.8\textwidth]{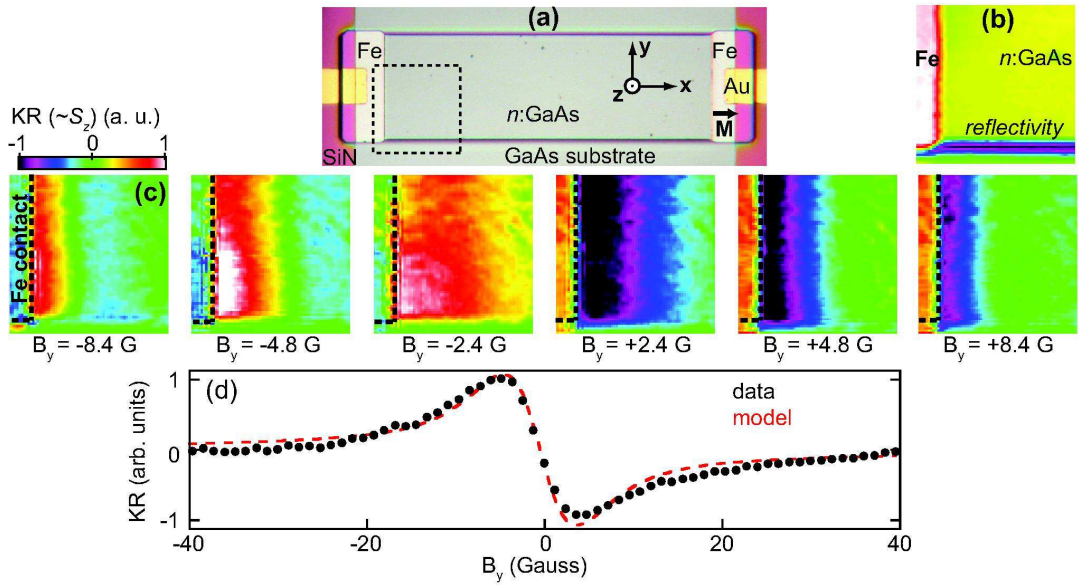}
\caption{(Color online) (a) A spin transport device having a 300
$\mu$m long \emph{n}:GaAs channel separating Fe/GaAs source and
drain contacts. The dotted square shows the $80 \times 80$ $\mu$m
region imaged. (b) An image of the reflected laser power. (c) Images
of electrical spin injection and transport in the \emph{n}:GaAs
channel ($V_b=0.4$ V, $I=92$ $\mu$A). Electrons are injected with
initial spin polarization $S_0 \parallel -\hat{x}$ and $B_y$ is
varied from -8.4 to +8.4 G (left to right), causing spins to precess
out of and into the page ($\pm \hat{z}$) respectively. (d) With the
probe laser positioned 4 $\mu$m from the Fe contact, measuring KR
($\propto S_z$) vs $B_y$ gives this ``Hanle curve". The dashed red
line is a simulation using $\tau_s$=125 ns, $v_d$=24000 cm/s, and
$D=10$ cm$^2$/s.} \label{fig4}
\end{figure*}

These Kerr images were acquired with the probe laser tuned to a
photon energy of 1.513 eV, just below the bandgap of the
\emph{n}:GaAs.  The optical KR that is due to the presence of
spin-polarized electrons in the \emph{n}:GaAs is strongly dependent
on photon energy, and its explicit dependence in \emph{this} device
is shown in Fig. 2(f). The exact shape of this curve varies from
device to device, and depends in part on the thickness and doping
density of the \emph{n}:GaAs layer. Further, once this
sample-specific and spin-dependent `fingerprint' is established,
spectral shifts of this curve provide a sensitive and quantitative
measure of strain (intentional or otherwise) in the sample
\cite{CrookerPRL}. The shape and sign of this curve can change if
the probe laser is positioned over other features on the device such
as the Fe contacts. For example, at this probe energy of 1.513 eV,
the spin polarized electrons that have diffused under the Fe
contacts in Figs. 2(b-d) lead to a KR of opposite sign (black regions
in the images; see color scale).  This can also be observed in Figs. 4 and 5.

Figure 3 shows measurements of this same device for the case of
electrical spin injection. Fig. 3(a) sketches the experiment,
wherein a voltage bias is applied across the two Fe fingers.
Spin-polarized electrons at the Fermi level of the Fe tunnel
through the thin Schottky barrier defined by the $n^+$:GaAs region
and into the \emph{n}:GaAs epilayer. The initial spin polarization
$S_0$ of these injected electrons is in-plane and antiparallel to
the Fe magnetization \textbf{M} (corresponding to majority spins in
Fe \cite{CrookerScience}).  A small magnetic field, also in-plane
but orthogonal to $S_0$, is used to precess these injected spins to
the out-of-plane direction ($\pm \hat{z}$) so that they can be
measured by the polar Kerr effect.  Figs. 3(b) and (c) show images
of the electrically injected spin polarization, where the injected
spins are tipped into the $+\hat{z}$ and $-\hat{z}$ direction by a
positive and negative in-plane magnetic field.

We confirm that these electrically-injected spins induce the same KR
spectral `fingerprint' as for the previous case of
optically-injected spins in this device. With the probe laser
positioned on the \emph{n}:GaAs near the Fe finger, the KR was
measured versus probe energy for both positive and negative in-plane
magnetic field. The red line in Fig. 3(d) shows the
\emph{difference} of these two curves, which eliminates any
field-independent birefringent offsets that can arise from
electrical modulation, and leaves behind only the signal that
depends on electron spin precession. This purely spin-dependent
signal agrees very well within an overall scale factor with the
previous KR signal resulting from optical spin injection (black
curve).

Figure 4(a) shows one of a later series of spin transport devices
having rectangular Fe/GaAs source and drain contacts at either end
of a long \emph{n}:GaAs channel. Studies of electrical spin
injection, accumulation and transport in these devices were reported
in Ref. \cite{CrookerScience}, and all-electrical detection of spin
accumulation was reported in Ref. \cite{XiaohuaPRL}. In Fig. 4 we
show the effect of in-plane magnetic fields $B_y$ on images of
electrically-injected spins.  We image an $80 \times 80$ $\mu$m
region that includes part of the Fe injection contact and the bottom
edge of the \emph{n}:GaAs channel (dotted square in Fig. 4(a)). A
reflectivity image (see Fig. 4b) clearly shows these features.  With
$V_b=0.4$ V, Fig. 4(c) shows a series of KR images of the
electrically-injected spins as $B_y$ is varied from -8.4 G to +8.4
G. Injected electrons, spin polarized initially along the $-\hat{x}$
direction, precess into the $+\hat{z}$ or $-\hat{z}$ direction when
$B_y$ is oriented along $-\hat{y}$ or $+\hat{y}$. These injected
electrons flow down the channel with average drift velocity $v_d$
that is the same in all the images. The drifting spins precess at a
rate proportional to $|B_y|$; thus, the spatial period of the
observed spin precession is short when $|B_y|$ is large.

\begin{figure}[tbp]
\includegraphics[width=.40\textwidth]{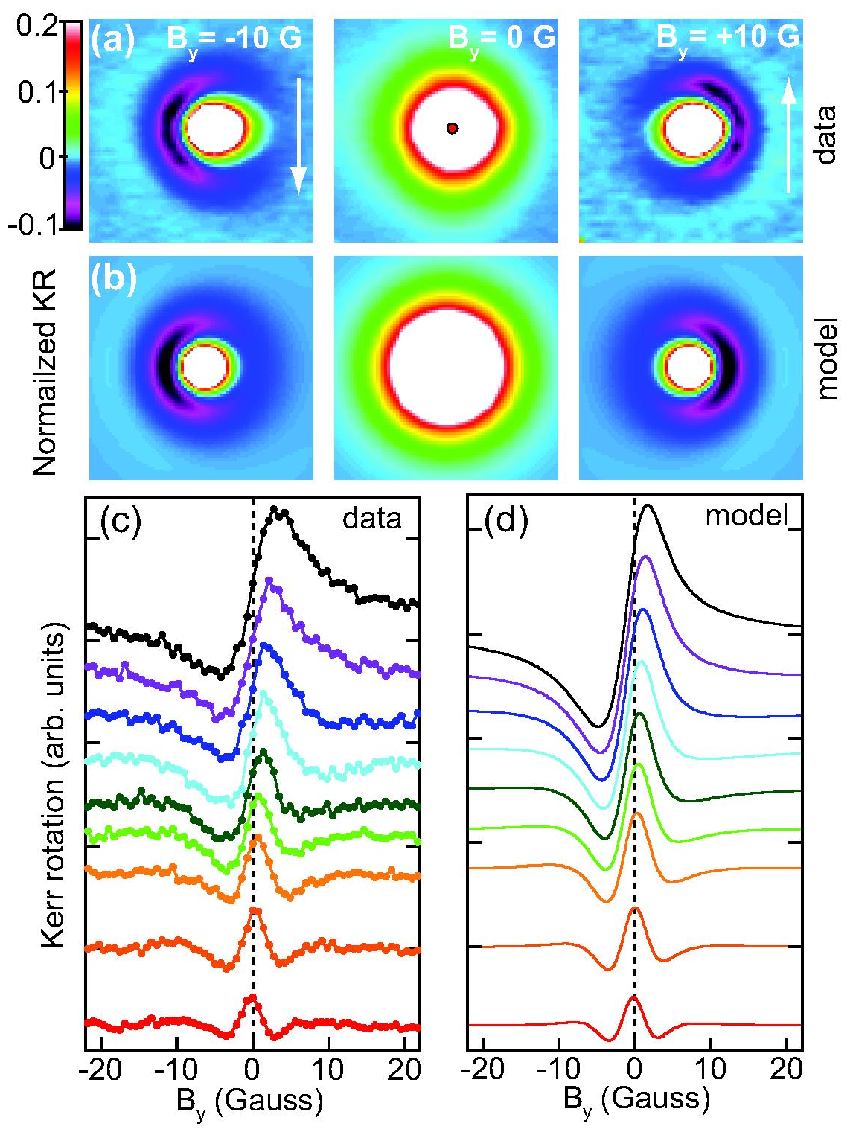}
\caption{(Color online)  Determining the presence and consequences
of residual off-diagonal strain $\epsilon_{xy}$ in the device. (a)
Three $80 \times 80$ $\mu$m images of optically-injected spin
polarization in the \emph{n}:GaAs channel, acquired with $B_y$ =
-10, 0, and +10 G respectively. The images are asymmetric due to the
combined influence of $B_y$ and the strain-induced `effective' field
$B_\epsilon$, which is oriented along $-\hat{y}$ or $+\hat{y}$ for
electrons diffusing to the left or to the right, respectively.  (b)
2D numerical simulations of spin diffusion using $\epsilon_{xy} =
0.8 \times 10^{-4}$, for $B_y$ = -10, 0, and +10 G. (c) Hanle curves
($S_z$ vs $B_y$) acquired at 2, 6, 10, 14, 18, 22, 26, 34, and 42
$\mu$m from the Fe/GaAs source contact for the case of electrical
spin injection ($V_b = 0.3$ V, $I=60$ $\mu$A). $\epsilon_{xy}$
shifts the peak of the Hanle curves to the left. (d) Modeling these
Hanle curves using $\epsilon_{xy} = 0.8 \times 10^{-4}$, $\tau_s =
125$ ns, $v_d= 18000$ cm/s, and $D=10$ cm$^2$/s.} \label{fig5}
\end{figure}

This series of images helps to make clear why, when the probe laser
is fixed at a point in the \emph{n}:GaAs channel and $S_z$ is
measured as an explicit function of $B_y$, we obtain ``Hanle curves"
having the characteristic antisymmetric lineshape shown in Fig.
4(d). The detailed structure of these Hanle curves (\emph{i.e.},
their amplitudes, half-widths, and oscillations) contains
considerable information about the dynamics of electron spin
transport in these devices including spin lifetime $\tau_s$,
diffusion constant $D$ and drift velocity $v_d$
\cite{CrookerScience, Hruska}. For the effectively one-dimensional
spin transport realized in this device, an analytic integral
solution to the spin drift-diffusion equations is readily derived
\cite{CrookerScience} and these Hanle curves can be accurately
modeled (see dotted red line).  We verify also that these curves
invert when the magnetization \textbf{M} of the Fe contacts is intentionally reversed (compare, \emph{e.g.}, with the Hanle curves in Fig. 5), and confirm that \textbf{M} is not affected by $B_y$.

Imaging studies also reveal a region of spin accumulation in the
\emph{n}:GaAs channel near the Fe drain contact.  Spin accumulation
in these devices results from spin-dependent transmission and
reflection of electrons at the Fe/GaAs tunnel barrier and was
studied in detail in Ref. \cite{CrookerScience}, and was also
investigated in forward-biased MnAs/GaAs structures by Stephens
\emph{et al} \cite{Stephens}.

Figure 5 shows how we detect the presence of off-diagonal strain, $\epsilon_{xy}$, in these devices,
and shows also how $\epsilon_{xy}$ manifests in spin transport
studies.  The device is the same as that shown in Fig. 4 and -- in
this case -- the strain was inadvertent, resulting most likely from
improper mounting and cooldown of the device. Figure 5(a) shows
images of spin-polarized electrons, optically injected in the middle
of the \emph{n}:GaAs channel, diffusing radially away from the point
of injection.  The applied magnetic field in the three images is
$B_y=-10$, 0, and +10 G respectively (see white arrows). The images
are clearly asymmetric in the presence of $B_y$, and this asymmetry
inverts when $B_y$ reverses. This asymmetry provides direct evidence
for the presence of off-diagonal strain in this device, and arises
from the asymmetric \emph{net} magnetic field `seen' by the
electrons, which are diffusing along all momentum directions
\textbf{k} in the \emph{x-y} plane. The net field is the vector sum
of both the applied magnetic field $B_y$ and a \textbf{k}-dependent
\emph{effective} magnetic field $B_\epsilon$ that is due to
spin-orbit coupling to strain \cite{OO, CrookerPRL}: $B_\epsilon
\propto \epsilon_{xy}(\sigma_y k_x - \sigma_x k_y)$. $B_\epsilon$
describes an effective field that is always in-plane and orthogonal
to \textbf{k}, and is oriented along $\pm \hat{y}$ for spins
diffusing to the right or left. When $B_y$ is negative (in Fig.
5(a)), electron spins diffusing to the left `see' a large net
magnetic field and precess (giving negative KR), while
spins diffusing to the right see little or no net field ($B_y$ and
$B_\epsilon$ oppose each other) and do not precess, resulting in an
asymmetric image. Carefully remounting the sample eliminated this
accidental strain, and subsequent images in the presence of $B_y$
revealed a symmetric annulus of negative KR, as expected. Other
methods to detect strain and its influence on electron spins have
also been demonstrated, for example, based on the shift of
photoluminescence Hanle curves with the device under electrical bias
\cite{Korenev}, or on time-resolved precession of flowing electrons
in zero magnetic field \cite{KatoNature}.

These asymmetric KR images can be modeled by numerically solving a
set of strain-dependent spin-drift-diffusion equations, derived in
Refs. \cite{CrookerPRL, Hruska}. Figure 5(b) shows modeled data
using known sample parameters and a small off-diagonal strain:
$\epsilon_{xy} = 0.8 \times 10^{-4}$. Note this strain is over two
orders of magnitude smaller than typical strains associated with,
for example, biaxial strain due to lattice-mismatched growth. These
images thus provide a sensitive diagnostic to quantify the presence
of $\epsilon_{xy}$ in these devices, particularly when $\tau_s$ is large.

Despite the small value of $\epsilon_{xy}$ inferred from the images
of Fig. 5(a), this strain manifests directly in studies of
electrically-injected spin transport.  Figure 5(c) shows Hanle
curves ($S_z$ versus $B_y$) acquired in the \emph{n}:GaAs channel of
this device, at increasing distances from the Fe/GaAs source
contact. Near the source contact (black curve, 2 $\mu$m away), $S_z$
is an odd function of $B_y$, as expected and as discussed above.
Moving down the \emph{n}:GaAs channel, the curves become narrower
(reflecting the increasing `age' of the measured electrons
\cite{CrookerScience}) and, more importantly, they shift to the
left.  At a distance of 42 $\mu$m from the source contact, $S_z$ has
become an even function $B_y$ (red curve). This shift is due to the
presence of $\epsilon_{xy}$ and its associated $B_\epsilon$, which
augments $+B_y$ for electrons flowing down the channel. Again, these
Hanle curves can be modeled by numerically solving the
spin-drift-diffusion equations in the presence of strain.  Figure
5(d) shows the modeled data, again using $\epsilon_{xy} = 0.8 \times
10^{-4}$.

It was also demonstrated in Ref. \cite{CrookerScience} that these
Fe/GaAs Schottky tunnel barriers can function as electrical spin
\emph{detectors} in addition to their role as spin injectors. To
demonstrate spin-dependent conductivity through a Fe/GaAs contact,
we use the experimental geometry sketched in Fig. 6(a). We optically
inject spin polarized electrons into the \emph{n}:GaAs channel using
the circularly polarized pump laser. By current-biasing the device,
we cause these spins to flow to and through the Fe/GaAs drain
contact. The spin polarization of this current at the drain contact
can be tipped parallel or antiparallel to the Fe magnetization
\textbf{M} using a small magnetic field $\pm B_y$.  We measure the
device conductance, $G$, as a function of $B_y$. This experiment is
the inverse of the Kerr-effect measurements described in the first
part of this paper: Instead of optically measuring the $\hat{z}$
component of drifting spins that are electrically injected along
$\pm \hat{x}$, here we \emph{electrically} measure the $\hat{x}$
component of drifting spins that are \emph{optically} injected along
$\pm \hat{z}$. The drift-diffusion equations apply equally, and
therefore $G(B_y)$ has the same characteristic antisymmetric ``Hanle
curve" shape.  Figure 6(b) shows the normalized conductance change
$\Delta G/G$ versus $B_y$ for spins that were optically injected 40
$\mu$m ``upstream" from the edge of the Fe/GaAs drain contact, for
varying pump powers. The conductance change between spins oriented
parallel or antiparallel to \textbf{M} is not large -- of order one
part in $10^5$ -- but the signal-to-noise ratio measured in this way
is nonetheless excellent. Lastly, Fig. 6(c) shows $\Delta G/G$
versus $B_y$ at three different current biases for spins optically
injected 25 $\mu$m from the drain. At low current the curves are
narrow (black), reflecting the long time required for spins to drift
from the point of injection to the drain contact. At high current
bias the spins drift quickly to the drain and the curve is
correspondingly much broader, as expected (blue curve). The data in
Fig. 6(c) are inverted compared to Fig. 6(b), reflecting the fact
that the magnetization \textbf{M} of the drain contact was reversed
between these two data sets. In this device, the conductance is
largest when the electron current flowing through the drain is spin
polarized parallel to \textbf{M}.

\begin{figure}[tbp]
\includegraphics[width=.40\textwidth]{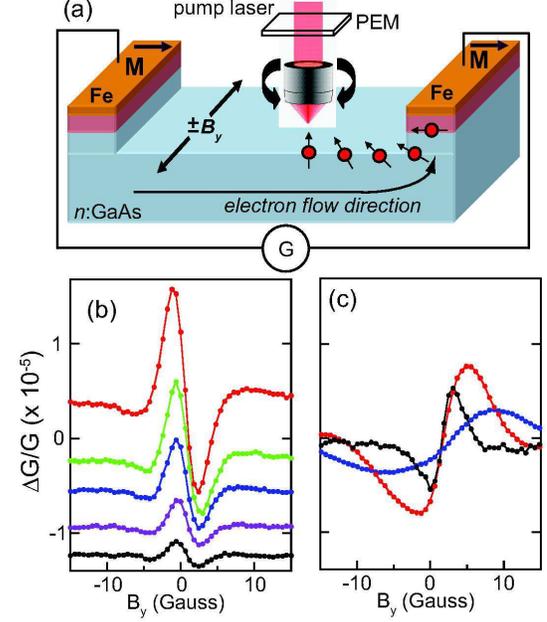}
\caption{(Color online)  Electrical detection of optically-injected
spin polarized currents.  (a) Schematic: The device is current
biased, and optically-injected spins (polarized initially along $\pm
\hat{z}$) flow to and through the Fe/GaAs drain contact. External
fields ($\pm B_y$) precess this spin-polarized current parallel or
antiparallel to \textbf{M}. The conductance $G$ is measured as a
function of $B_y$. (b) Normalized conductance change $\Delta G/G$ vs
$B_y$ for increasing pump laser power (10, 20, 50, 100, and 200
$\mu$W, from bottom to top). The spins are injected 40 $\mu$m from
the edge of the drain contact, and $V_b$=360 mV. (c) $\Delta G/G$ vs
$B_y$ for spins injected 25 $\mu$m from the drain contact, for
device biases of 160 mV (black), 440 mV (red), and 725 mV (blue).}
\label{fig6}
\end{figure}
In conclusion we have discussed methods, based on the
magneto-optical Kerr effect, to study and image both optically- and
electrically-injected spin polarizations in GaAs. These measurements
help to characterize spin transport phenomena in lateral Fe/GaAs
devices and suggest routes for all-electrical studies of
spin-dependent transport in hybrid ferromagnet-semiconductor
structures. This work was supported by the DARPA SpinS and Los
Alamos LDRD programs, the NSF MRSEC program under grant DMR
02-12302, the Office of Naval Research, and the Minnesota
Nanofabrication Center, which is supported by the NSF NNIN program.

%---------------------------------------------------------------
%***************************************************************
%---------------------------------------------------------------

%\newpage

\begin{references}

\bibitem{Johnson}M. Johnson and R. H. Silsbee, Phys. Rev. Lett. \textbf{55}, 1790 (1985).

\bibitem{Jedema}F. J. Jedema, H. B. Heersche, A. T. Filip, J. J. A. Baselmans, B. J. van Wees, Nature \textbf{416}, 713 (2002).

\bibitem{Valenzuela}S. O. Valenzuela and M. Tinkham, Nature \textbf{442}, 177 (2006).

\bibitem{OO}G. E. Pikus and A. N. Titkov, in \emph{Optical Orientation}, F. Meier and B. P. Zakharchenya, Eds. (North-Holland, Amsterdam,1984), pp. 73-131.

\bibitem{CrookerPRL}S. A. Crooker and D. L. Smith, Phys. Rev. Lett.  \textbf{94}, 236601 (2005).

\bibitem{CrookerScience}S. A. Crooker, M. Furis, X. Lou, C. Adelmann, D. L. Smith, C. J. Palmstr{\o}m, and P. A. Crowell, Science \textbf{309}, 2191 (2005).

\bibitem{Hruska}M. Hru\v{s}ka, S. Kos, S. A. Crooker, A. Saxena, and
D. L. Smith, Phys. Rev. B \textbf{73}, 075306 (2006).

\bibitem{Furis}M. Furis, D. L. Smith, J. L. Reno, and S. A. Crooker,
Appl. Phys. Lett. \textbf{89}, 102102 (2006).

\bibitem{XiaohuaPRL}X. Lou, C. Adelmann, M. Furis, S. A. Crooker, C. J. Palmstr{\o}m,
and P. A. Crowell, Phys. Rev. Lett. \textbf{96}, 176603 (2006).

\bibitem{Kikkawa}J. M. Kikkawa and D. D. Awschalom, Nature \textbf{397}, 139 (1999).

\bibitem{DzhioevPRB}R. I. Dzhioev \emph{et al.}, Phys. Rev. B \textbf{66}, 245204 (2002).

\bibitem{Hanbicki}A. T. Hanbicki \emph{et al}, Appl. Phys. Lett. \textbf{82}, 4092 (2003).

\bibitem{Stephens}J. Stephens, J. Berezovsky, J. P. McGuire, L. J. Sham, A. C. Gossard, D. D. Awschalom, Phys. Rev. Lett. \textbf{93}, 097602 (2004).

\bibitem{Korenev} V. K. Kalevich and V. L. Korenev, JETP Lett. \textbf{52}, 230 (1990).

\bibitem{KatoNature}Y. Kato, R. C. Myers, A. C. Gossard, and D. D. Awschalom, Nature \textbf{427}, 50 (2004).

%\bibitem{CrookerIJQE}S. A. Crooker, D. D. Awschalom, and N. Samarth, IEEE J. of Sel. Top. in Quantum Electronics \textbf{1}, 1082 (1995).

%\bibitem{Adelmann}C. Adelmann, X. Lou, J. Strand, C. J. Palmstr{\o}m, and P. A. Crowell, Phys. Rev. B \textbf{71}, 121301 (2005).

\end{references}
\end{document}